\documentclass[%
 reprint,
 amsmath,amssymb,
 aps,
 showkeys,
]{revtex4-2}

\usepackage{graphicx}% Include figure files
\usepackage{dcolumn}% Align table columns on decimal point
\usepackage{bm}% bold math
\usepackage[colorlinks=true,allcolors=blue]{hyperref}
\usepackage[dvipsnames]{xcolor}
\begin{document}

\preprint{APS/123-QED}

\title{The Role of Phase and Spatial Modes in Wave-Induced Plasma Transport}% Force line breaks with \\

\author{L. F. B. Souza}
\email{luisfernando.bernardi1998@gmail.com}
\affiliation{Institute of Physics, University of São Paulo, São Paulo, 13506-900, SP, Brazil}
\affiliation{Aix-Marseille Université, CNRS, UMR 7345 PIIM, F-13397, Marseille cedex 13, France}

\author{Y. Elskens}
\affiliation{Aix-Marseille Université, CNRS, UMR 7345 PIIM, F-13397, Marseille cedex 13, France}

\author{R. Egydio de Carvalho}
\thanks{Senior professor}
\affiliation{Department of Statistics, Applied Mathematics and Computer Science, São Paulo State University, Rio Claro, 13506-900, SP, Brazil}

\author{I. L. Caldas}
\affiliation{Institute of Physics, University of São Paulo, São Paulo, 13506-900, SP, Brazil}

\begin{abstract}
We derive a two-dimensional symplectic map for particle motion at the plasma edge by modeling the electrostatic potential as a superposition of integer spatial harmonics with relative phase shift, then reduce it to a two-wave model to study the transport dependence on the perturbation amplitudes, relative phase, and spatial-mode choice. Using particle transmissivity as a confinement criterion, identical-mode pairs exhibit phase-controlled behavior: anti-phase waves produce destructive interference and strong confinement while in-phase waves add constructively and drive chaotic transport. Mode-mismatched pairs produce richer phase-space structure with higher-order resonances and sticky regions; the transmissivity boundaries become geometrically complex. Box-counting dimensions quantify this: integer dimension smooth boundaries for identical modes versus non-integer fractal-like dimension for distinct modes, demonstrating that phase and spectral content of waves jointly determine whether interference suppresses or promotes transport.
\end{abstract}


\maketitle

%\tableofcontents

\section{Introduction}

Turbulence-driven particle transport in magnetically confined plasmas is a well-known and persistent challenge \cite{RCWolf2003, hazeltine2013plasma}. This phenomenon significantly hinders plasma confinement by causing undesirable particle losses \cite{WHorton1985, horton2012turbulent}. The turbulence of toroidal plasmas in machines like tokamaks is strongly affected by a large number of instabilities \cite{post1986introduction}. One of the most significant phenomena in magnetized plasmas is the occurrence of drift instabilities, which arise in the presence of sharp density gradients within the plasma column. These instabilities are frequently observed in plasmas with nonuniform density, where equilibrium is maintained by a strong magnetic field, giving them a universal nature.

In this first attempt to include multiple coherent waves within this reduced Hamiltonian framework, we intentionally treat wave amplitudes, spatial modes, and relative phases as prescribed (time-independent) parameters in order to isolate and quantify how phase offsets and spectral mode content directly control single-particle transport. In realistic plasmas, however, nonlinear processes—particularly modulational interactions known to generate phase correlations and redistribute spectral energy\cite{Benkadda2011,Vladimirov1995}, can dynamically modify these quantities. Incorporating such self-consistent nonlinear wave–wave dynamics would require a substantially different theoretical framework and lies beyond the scope of the present work, which is focused on identifying the transport consequences of given spectral and phase configurations.

At low frequencies, magnetic perturbations have a negligible effect, and drift waves can be considered electrostatic. In this case, the electric field aligns with the propagation direction, satisfying the condition $\nabla \times \mathbf{E} = 0$, which allows \cite{WHorton1985} the definition of a potential such that $\mathbf{E} = -\nabla \tilde{\phi}$. The interaction between the electrostatic field of the drift wave and the dominant magnetic field leads to particle motion governed by the $\mathbf{E} \times \mathbf{B}$ drift. Moreover, turbulence in tokamak plasmas is primarily influenced by $\mathbf{E} \times \mathbf{B}$ drift effects. At the plasma edge, the $\mathbf{E}\times \mathbf{B}$ drift in the particle motion causes changes in the transport coefficients, this phenomenon is known as anomalous transport \cite{weiland2019drift}. Over the years, various control mechanisms have been proposed to reduce this anomalous transport, both theoretically and experimentally. These include modifying the radial component of the electric field through the application of an external electric potential to create transport barriers and reduce transport coefficients \cite{burrell1997effects, connor2004review, garbet2004physics}.

 The model introduced by Horton \cite{WHorton1985}, also known as the drift wave model, exhibits a Hamiltonian structure, specifically a non-autonomous Hamiltonian with one and a half degrees of freedom.  This model inherently generates chaotic dynamics through deterministic electrostatic fluctuations. Chaotic trajectories can emerge, extending from the outer edge of the plasma column to the tokamak wall. This behavior makes the model a valuable framework for studying the influence of $\mathbf{E} \times \mathbf{B}$ drift on particle loss within a Hamiltonian formulation. By introducing action-angle variables on an appropriately selected Poincaré surface, particle dynamics can be described through a symplectic map \cite{horton1998}. In this work, the map equations incorporate radial profiles of equilibrium quantities associated with electromagnetic fields, which are fundamental in present tokamak configurations, allowing for a more realistic description of particle dynamics.

Previous studies \cite{marcus2019influence, rosalem2014, miskane2000anomalous, souza2022, osorio2023, osorio2021} have shown that non-monotonic equilibrium radial confinement profiles induce topological changes in phase space. These properties, particularly in particle confinement studies, manifest in the transport coefficients, which measure the flux of particles or energy across the system. Moreover, the values of these coefficients depend on the presence and structure of transport barriers.

A key feature in this context is the shearless transport barrier (STB), which strongly limits particle transport and improves confinement. As the perturbation intensity increases, chaotic layers expand and begin to surround island chains and nearby invariant tori, including the shearless torus. With further increasing of the perturbation, these invariant tori are progressively destroyed. Nonetheless, the destroyed shearless torus may continue to influence dynamics via stickiness: chaotic trajectories linger near the remnant structure for long times, producing prolonged escape times and effectively reducing transport fluxes. \cite{nascimento2005plasma, marcus2008}

The motion of a particle in the field of many waves is a known process in Hamiltonian physics. Previous studies have shown the effect of waves on particle transport \cite{elskens2010nonquasilinear, kono2010nonlinear}. The proposed symplectic drift wave map assumes a dominant spatial mode wave potential as the source of non-integrability (perturbation) \cite{horton1998}. Here, we consider a superposition of multiple coherent waves, with the integer multiple spatial mode but with a constant phase difference \(\nu\) between them. As a result, the perturbation in the system depends not only on the amplitude of the waves but also on their relative phases. The inclusion of a multiple waves modifies the single wave map by introducing multiple periodic perturbative terms in the action coordinate. The one-wave map shares similarities with the standard non-twist map \cite{diego1996}, as both are two-dimensional, non-twist area-preserving maps with a single periodic perturbation term. The multiple map, on the other hand, is analogous to the Labyrinthic map \cite{martins2010} and extended non-twist map \cite{mugnaine2020, Souza2024}, all of which involve more than one periodic perturbative terms.

In this paper we derive a two-dimensional symplectic map from a superposition of waves with integer spatial modes and phase offsets, with the aim of investigate how perturbation amplitude, relative phase and the spatial spectrum shape particle transport in the plasma edge. We focus on a reduced two-wave model and compare two representative scenarios: waves with the same spatial mode and waves with distinct spatial modes. By scanning parameters and evaluating transmissivity, we identify how phase and mode choice either suppress or enhance transport, and we show that mode-mismatched pairs produce significantly richer dynamics—additional resonances, sticky regions and complex transport boundaries—than identical-mode pairs. Finally, we quantify these differences using the box-counting method to characterize the geometry of the transport/no-transport boundary, providing an objective measure that links the observed dynamical complexity to the fractal nature of parameter-space boundaries.

The paper is organized as follows. In Section \ref{section2}, we derive the multiple-wave map. In Section \ref{section3}, we present the transport study for identical spatial modes. Section \ref{section4} compares the dynamics for distinct spatial modes. Section \ref{section5} reports box-counting dimension results for both cases. Section \ref{conclusion} summarizes our conclusions.

\section{Drift wave model}\label{section2}

To investigate the influence of electrostatic oscillations on particle transport, we consider a dynamical system based on the differential equations introduced by Horton \cite{WHorton1985}, which describe a magnetically confined plasma in a tokamak. We consider the motion of a test particle in a tokamak plasma with a large aspect ratio (\(a/R \ll 1\)), where \(a\) and \(R\) are the minor and major radii, respectively. The guiding center of the particle follows the magnetic field lines, \(\mathbf{B}(\mathbf{x})\), with a parallel velocity \(\upsilon_{||}(\mathbf{x})\) along \(\mathbf{B}\), while also drifting due to the \(\mathbf{E \times B}\) velocity. The equations governing the particle motion are given by
\begin{equation}\label{Eq1}
     \frac{\mathrm{d}\mathbf{x}}{\mathrm{d}t} = v_{||}\frac{\mathbf{B}}{B} + \mathbf{v_{E \times B}}, \quad \mathbf{v_{E \times B}} = \frac{\mathbf{E \times B}}{B^{2}},
\end{equation}
where $\mathbf{x} = (r, \theta, \varphi)$ are toroidal coordinates. The magnetic configuration considered is
\begin{equation}\label{Eq2}
    \mathbf{B}(r) = B_{\theta}(r)\hat{e}_{\theta} + B_{\varphi}(r)\hat{e}_{\varphi}
\end{equation}
with  $B \approx B_{\varphi} \gg B_{\theta}$ and $B \approx B_0$, where $B_0$ is constant. In addition, the electric field is treated as a curl-free vector field, $\nabla \times \mathbf{E} = 0$. Under equilibrium conditions, it is fully determined by its radial component, $E_r(r) \hat{e}_r$. Outside equilibrium, a simplified drift wave transport model is employed, accounting for electrostatic potential fluctuations, $\tilde{\phi}(t, \mathbf{x})$, which are represented by multiple spatial modes along with harmonics of the lowest dominant angular frequency, $\omega_0$, in the drift wave spectrum, described by the Fourier expansion
\begin{subequations}\label{Eq3}
\begin{align}
    & \mathbf{E}(t, r,\theta, \varphi) = E_r(r) \hat{e}_r - \nabla  \tilde{\phi}(t, \theta,\varphi), \\
    & \tilde{\phi}(t, \theta,\varphi) = \sum_{l,m, n} \phi_{l,m,n} \cos (m\theta - l\varphi - n\omega_0 t + \nu_{l,m,n}).
\end{align}
\end{subequations}
where $m,l$ are the spatial modes, $n$ is the temporal modes, and $\phi_{l,m,n}$,$\nu_{l,m,n}$ are the amplitude and the phase of each mode. For simplicity, we consider the Fourier coefficients constant in the region of interest $(\partial_r \tilde{\phi} = 0)$. We focus on a dominant spatial mode with $m = M$ and $l = L$, and consider the electrostatic potential as a superposition of infinity waves with integer multiple  of the dominant spatial mode with a relative phase difference, so the potential assumes the following form:
\begin{equation}\label{Eq4}
    \tilde{\phi}  = \sum_k \sum_{n = -\infty}^{\infty} \phi_{n}^{k} \cos (\eta_k M\theta - \eta_k L\varphi - n\omega_0 t + \nu_{n}^{k})
\end{equation}
where $\phi_{n}^{k}$ is the amplitude of each wave. Transforming to action $I \equiv (r/a)^2$  and angle $\Psi \equiv M\theta - L\varphi$ variables, the electrostatic potential becomes
\begin{equation}\label{Eq5}
    \tilde{\phi}(\Psi, t) =  \sum_k \sum_{n = -\infty}^{\infty} \phi_{n}^{k} \cos (\eta_k \Psi - n\omega_0 t + \nu_{n}^{k}).
\end{equation}
Considering correlated harmonics $(\nu_{n}^{k} = \nu_k)$ and using trigonometric identities, we rewrite the potential as
\begin{equation}\label{Eq6}
       \tilde{\phi}(\Psi, t) =   2\pi \sum_k \phi_{k} \cos(\eta_k \Psi + \nu_{k})\sum_n \delta(\omega_0 t - 2\pi n).
\end{equation}
The mathematical steps from Eq.\,\eqref{Eq5} to Eq.\,\eqref{Eq6} are in Appendix~\ref{Appendix2}. In the new coordinates, the drift equations become
\begin{subequations}\label{Eq7}
    \begin{align}
        \frac{\mathrm{d}I}{\mathrm{d}t} &= \dfrac{4\pi M}{a^{2}B} \sum_k \phi_{k} \eta_k \sin(\eta_k \Psi + \nu_{k})\sum_n \delta(\omega_0 t - 2\pi n), \\
        \frac{\mathrm{d}\Psi}{\mathrm{d}t} &= \dfrac{\upsilon_{||}(I)}{R}\left[\dfrac{M}{q(I)} - L\right] - \dfrac{ME_r(I)}{aB\sqrt{I}}.
    \end{align}
\end{subequations}
where $q(r) = \frac{rB_{\varphi}}{RB_{\theta}}$ is the safety factor profile. 
From Eqs.\,\eqref{Eq7}, one obtains
\begin{equation}\label{Eq8}
    \frac{\partial \dot{I}}{\partial I} + \frac{\partial \dot{\Psi}}{\partial \Psi} = 0,
\end{equation}
satisfying the Liouville theorem. Moreover, the system admits a Hamiltonian description: there exists a function \(H(t,\Psi,I)\) generating the dynamics through the canonical equations
\begin{equation}\label{Eq9}
    \dot{\Psi} = \frac{\partial H}{\partial I}, \qquad \dot{I} = -\frac{\partial H}{\partial \Psi}.
\end{equation}
The Hamiltonian function can be decomposed into an integrable component, \( H_0(I) \), and a perturbative term, \( H_1(t, \Psi) \), such that 
\begin{equation}\label{Eq10}
H(t, \Psi, I) = H_0(I) + H_1(t, \Psi). 
\end{equation}
Here, the integrable part is defined as
\begin{subequations} \label{Eq11}
\begin{align}
H_0(I) &= \int^{I} \frac{\mathrm{d}\Psi}{\mathrm{d}t} \, \mathrm{d}I', 
\end{align} 
while the perturbative term takes the form 
\begin{align}
H_1(t, \Psi) &= 2M\tilde{\phi}(t, \Psi). 
\end{align} 
\end{subequations} 
Thus, the variables \( (\Psi, I) \) serve as a set of canonical coordinates for the system. The perturbation term, Eq.\,\eqref{Eq6}, consists of periodic pulses with period $T = 2\pi/\omega_0$. Assuming the amplitudes $\phi_{n}^{k}$ and phase $\nu_{n}^{k}$ remain constant between pulses $(n, n +1)$, the drift wave map is obtained as
\begin{subequations}\label{Eq12}
    \begin{align}
        I_{n+1} &= I_n + \sum_k \alpha_k \sin(\eta_k \Psi_n + \nu_k),  \label{Eq10a}\\
        \Psi_{n+1} &= \Psi_n + g(I_{n+1}),
    \end{align}
\end{subequations}
with the function
\begin{equation}\label{Eq13}
    g(I) = \dfrac{\upsilon_{||}(I)}{\omega R}\left(\dfrac{M}{q(I)} - L\right) - \dfrac{ME_r(I)}{aB\omega \sqrt{I}}.
\end{equation}
The control parameters are
\begin{equation}\label{Eq14}
    \alpha_k = \dfrac{4\pi M}{a^{2}\omega B}\phi_{k} \eta_k.
\end{equation}

The dynamical system defined by Eq.~\eqref{Eq12} is a two-dimensional symplectic map expressed in action-angle coordinates. It models the motion of a particle in the field of multiple waves with integer wavenumbers, incorporating a phase difference between them. This map serves as a framework for studying how the interaction of these waves, and specifically their relative phase, influences particle transport and the onset of chaos.

The control parameters $\alpha_k$ are directly related to the wave amplitudes $\phi_{k}$ and the multiple integer mode $\eta_{k}$, as shown in \eqref{Eq14}. The function $g(I)$ in Eq.\,\eqref{Eq13} governs the evolution of the angle coordinate $\Psi$ and corresponds to the twist function in the context of Hamiltonian dynamics. Its behavior is determined by the choice of equilibrium profiles $(E_r, v_{||}, q)$, which influence the map’s properties and lead to topological rearrangements in phase space. In particular, non-monotonic radial profiles can result in the violation of the twist condition, giving rise to one or more STBs \cite{marcus2019influence, rosalem2014, miskane2000anomalous, osorio2021}.

\section{Profiles adopted}\label{section3}

We take into account the plasma profiles and parameters for the tokamak TCABR, mainly, the radial equilibrium electric field \cite{marcus2008}, $E_r(I)$, the parallel velocity \cite{severo2009}, $\upsilon_{||}(I)$, and the safety factor \cite{rosalem2016drift}, $q(I)$, which are specified in Eqs.\,\eqref{Eq15}, \eqref{Eq16}, and \eqref{Eq17}, respectively.  Such profiles are commonly observed in tokamaks operating in the H-mode (High-confinement mode) regime \cite{asdex1989h, grenfell2018h}. In this regime, a pedestal structure forms near the plasma edge, affecting the density, pressure, and temperature profiles \cite{wagner1982regime, connor2000review}. This structural change influences the characteristic radial electric field profile \cite{grenfell2018h, viezzer2013high, ida1990edge}.

For TCABR, the minor and major plasma radii are $a = 0.18$\,m and $R = 0.61$\,m, respectively, the tokamak minor radius $b = 0.21$\,m, and the toroidal magnetic field $1.2$\,T.  We used the following expressions, compatible with profiles measured in the TCABR tokamak:

\begin{equation}\label{Eq15}
E_r(I)= e_1I + e_2\sqrt{I} + e_3,
\end{equation}
\begin{equation}\label{Eq16}
    \upsilon_{||}(I) = \upsilon_1 + \upsilon_2\tanh{(\upsilon_3I + \upsilon_4)}, 
\end{equation}
\begin{equation}\label{Eq17}
    q (I)= 
    \begin{cases}
        q_0 + \left(q_a - q_0\right)I, \quad& I\leq 1\\
        q_a I, \quad& I> 1
    \end{cases}
    \end{equation}  
where the normalized values of the various coefficients are in Table\,\ref{table1}.

\begin{table}
    \centering
    \begin{tabular}{c@{\hspace{0.75cm}} c@{\hspace{0.75cm}} | c@{\hspace{0.75cm}} c@{\hspace{0.75cm}} | c@{\hspace{0.75cm}} c}
    \hline
    \hline
        $e_1$ & $10.7$ & $\upsilon_2$ & $17.47$ & $q_a$ & $3.65$ \\
      $e_2$  & $-15.8$ & $\upsilon_3$ & $10.1$ & $M$ & $16$ \\
      $e_3$ & $4.13$ & $\upsilon_4$ & $-9.00$ & $L$ & $4$ \\
      $\upsilon_1$ & $-9.867$ & $q_0$ & $1.00$ & $\omega$ & $16.36$ \\
      \hline
      \hline
    \end{tabular}
    \caption{Normalized parameter values for the two-dimensional drift map.}
    \label{table1}
\end{table}

 As described in Eq.\,\eqref{Eq13}, the function $g$ governs the evolution of the angle coordinate $\Psi$, acting as the twist function of the map. This function exhibits a maximum value, which marks the point where the twist condition, corresponding to the non-degeneracy of the Hamiltonian function, is violated. In the next section, we will show that this function is directly related to a key quantity in the orbit dynamics: the winding number (or rotation number). The emergence of a non-monotonic profile in $g(I)$ is precisely what characterizes the formation of a STB.

\section{Same spatial mode two-wave transport}\label{section4}
\begin{figure*}
    \centering
    \includegraphics[width=0.85\linewidth]{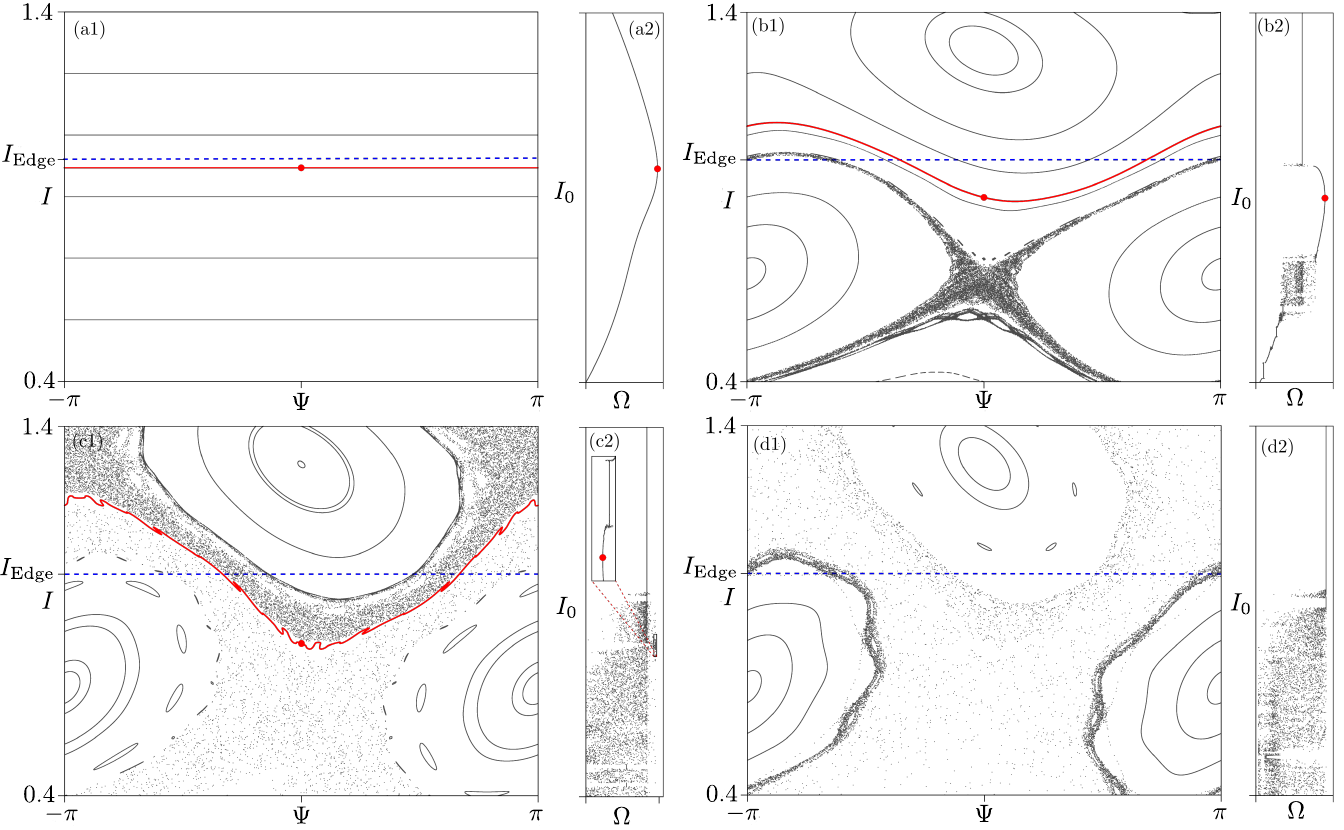}
    \caption{Evolution of phase space and corresponding winding number profiles for a single wave ($\alpha_2 = 0$) with increasing amplitude $\alpha_1$. Left column (a1-d1): Phase portraits for $\alpha_1 = 0, 0.09, 0.12, 0.18$. Right column (a2-d2): The associated winding number $\Omega(I)$. In all panels, the shearless curve and its corresponding extremum in the winding profile are highlighted in red.}
    \label{fig1}
\end{figure*}

\begin{figure*}[htbp]
    \centering
    \includegraphics[width=0.75\linewidth]{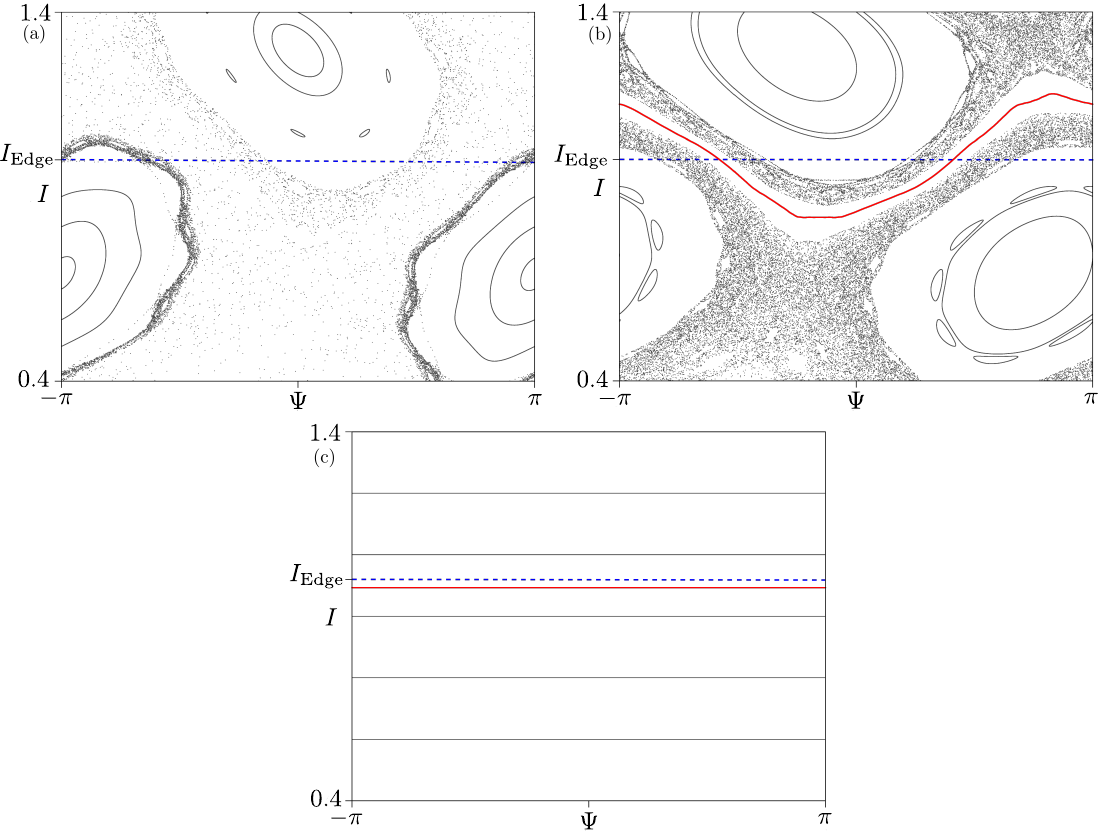}
    \caption{Phase space portraits in action-angle coordinates $(I, \Psi)$ for a fixed amplitude $\alpha_{1,2} = 0.09$, demonstrating the impact of wave interference. (a) Constructive interference ($\nu = 0$), (b) Intermediate phase ($\nu = \pi/2$), (c) Destructive interference ($\nu = \pi$). The STB is highlighted in red in each case.}
    \label{fig2}
\end{figure*}
\begin{figure*}
    \centering
    \includegraphics[width=0.85\linewidth]{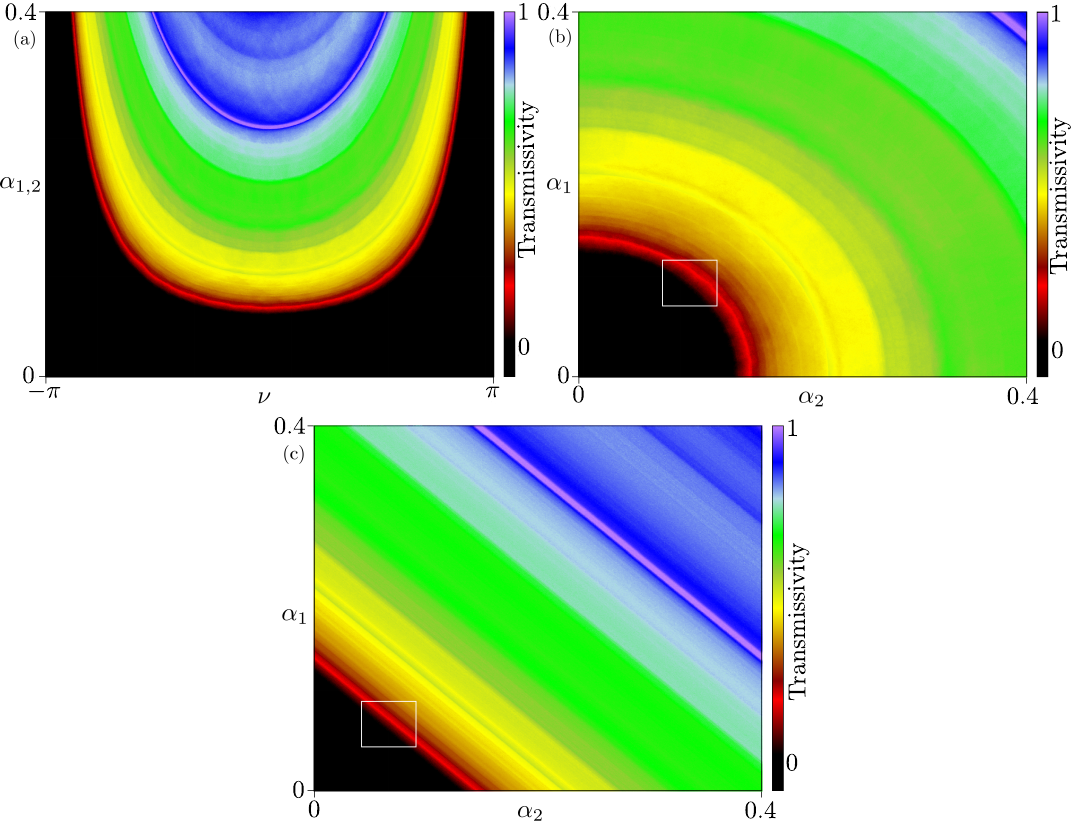}
   \caption{For $\eta_{1,2} = 1$ the transmissivity across different parameter spaces. (a) Phase-amplitude space $(\nu, \alpha_{1,2})$. (b, c) Amplitude-amplitude space $(\alpha_2, \alpha_1)$ for a fixed phase difference of (b) $\nu = 0$ and (c) $\nu = \pi/2$. In all panels, black denotes parameter combinations resulting in no transport.}
    \label{fig3}
\end{figure*}
We begin by reducing the infinitely many wave map \eqref{Eq12} to a two-wave system where both waves share the same dominant spatial mode $(M = 16, L = 4)$ but maintain a phase difference $\nu$. This yields the simplified symplectic map:

\begin{subequations}\label{Eq18}
\begin{align}
I_{n+1} &= I_n + \alpha_1 \sin(\Psi_n) + \alpha_2 \sin(\Psi_n + \nu), \\
\Psi_{n+1} &= \Psi_n + g(I_{n+1}),
\end{align}
\end{subequations}
where the control parameters $\alpha_{1,2}$ are given by Eq.\,\eqref{Eq14}. For this analysis, we set the spatial mode coupling to $\eta_1 = \eta_2 = 1$. Physically, having equal $\eta$ means the two perturbations share the same spatial harmonic (they are integer multiples of the same dominant spatial mode): their poloidal/toroidal spatial structure is identical and so they interfere directly in the action coordinate $I$. Appendix~\ref{Appendix3} demonstrates that the system can be reduced to one perturbation term map for two waves sharing the same spatial mode.

We analyze particle confinement by exploring the $(\Psi, I)$ phase space as a function of the perturbation amplitudes $\alpha_{1,2}$ and the phase difference $\nu$. For non-zero perturbations ($\alpha_{1,2} \neq 0$), the dynamics features a complex interplay between chaotic trajectories and regular orbits.

A key tool for distinguishing orbit types is the winding number $\Omega$, which quantifies the average rotational frequency of trajectories:
\begin{equation}\label{Eq19}
\Omega = \lim_{n \to \infty} \frac{\Psi_n - \Psi_0}{n},
\end{equation}
where $\Psi_0$ is the initial phase. If the orbit is chaotic the limit doesn't converge, while for regular orbits it converges. Orbits which the winding number converge to irrational values are quasiperiodic and form transport barriers that inhibit chaotic diffusion, thereby confining particle motion and enhancing plasma confinement.

Figure~\ref{fig1} illustrates this phenomenology for a single wave ($\alpha_2 = 0$). Panel (a1) shows the integrable case ($\alpha_1 = 0$), consisting only of invariant curves and periodic orbits. The red curve denotes the shearless curve, identified by an extremum point in the winding number profile in (a2). The blue dotted line represents the plasma edge ($I_{\mathrm{edge}} = 1.0$).

Introducing perturbation ($\alpha_1 > 0$) in panel (b1) breaks integrability, generating chaotic orbits around hyperbolic fixed points. These chaotic regions coexist with stable islands and, most importantly, invariant curves. The quasiperiodic orbits forming these curves act as transport barriers, visibly constraining the chaotic orbits to access the totality of the phase space \cite{meiss1992, Zaslavsky2000, diego1996, meiss2015thirty}. The shearless curve is highlighted in red.

Panel (c1) demonstrates a critical nuance: even at higher perturbation, the shearless curve can persist. However, due to its specific topology, it does not form a total barrier, since part of the curve is outside the plasma $(I > 1)$, then some particle loss can still occur. This shows that the mere presence of a shearless curve is an insufficient criterion for global confinement, necessitating a more comprehensive transport measure.

Finally, for a sufficiently large perturbation in panel (d1) ($\alpha_1 = 0.18$), all invariant structures, outside the islands, are destroyed. With no transport barriers remaining, chaotic orbits permeate the phase space, allowing particles to freely reach the tokamak wall. The range of normalized perturbation amplitudes examined here (e.g. $\alpha$ up to $\sim 0.18$) is chosen to be within experimentally relevant values for the TCABR edge when mapped from measured floating-potential/electric-field fluctuations to our normalization; comparable fluctuation amplitudes and biasing-induced variations have been reported in TCABR experiments. \cite{nascimento2005plasma,guimaraes2008electrostatic,kuznetsov2012long}.

We now introduce a second wave to explore the interplay between wave interference and transport. Figure~\ref{fig2} presents phase spaces for fixed wave amplitudes $\alpha_{1} = \alpha_{2} = 0.09$ under different phase shifts $\nu$.

Panel (a) ($\nu = 0$) shows a phase space identical to the single-wave case in Fig.\,\ref{fig1}(d1). This equivalence arises from constructive interference; for in-phase waves ($\nu = 0$), their amplitudes add linearly according to Eq.\,\eqref{Eq14}, making the system with $(\alpha_{1}, \alpha_{2}) = (0.09, 0.09)$ dynamically equivalent to a single wave with $\alpha_1 = 0.18$.

Panel (b) ($\nu = \pi/2$) reveals a more strongly perturbed scenario compared to Fig.\,\ref{fig1}(b1), characterized by expanded chaotic regions and fewer invariant curves.

Panel (c) ($\nu = \pi$) shows an integrable system, a result of perfect destructive interference. When completely out of phase ($\nu = \pm \pi$), the waves cancel each other's drive in the action coordinate $I$ according to Eq.\,\eqref{Eq14}, resulting in an integrable system where $I$ remains constant. This general behavior holds for any $\alpha_{1} = \alpha_{2}$ when $\nu = \pm \pi$.

A notable observation is the lower density of chaotic orbits in Fig.\,\ref{fig2}(a) compared to (b), despite identical initial conditions and iteration counts. This occurs because chaotic orbits in (a) more rapidly reach the region $I \le 0$, where the map is undefined, causing them to terminate earlier and reducing their visible density.

To systematically evaluate confinement across parameter regimes, we introduce the transmissivity metric. We consider an ensemble of $K_{\mathrm{IC}}$ initial conditions below the plasma edge ($I_0 = 0.75$) and iterate them until they either reach an upper threshold $I_{\mathrm{up}}= 1.25$ or a maximum number of steps $N_{\mathrm{max}} = 5\times 10^{3}$. The transmissivity is the fraction of orbits that escape.

A transmissivity of zero indicates a total transport barrier near the edge region prevents escape. A value between zero and one indicates partial barriers and sticky dynamics, while a value of one signifies no confinement.

Figure~\ref{fig3} presents transmissivity maps across key parameter spaces, generated from a $1600 \times 1600$ grid with $K_{\mathrm{IC}} = 1000$. The color represents the transmissivity: black, transitioning to red and blue for increasing escape fractions.

Panel (a) shows the space $(\nu, \alpha_{1,2})$. Regions near $\nu = \pm\pi$ exhibit zero transmissivity (black) even for large $\alpha_{1,2}$, confirming that destructive interference promotes integrability and strong confinement. In contrast, near $\nu = 0$ (constructive interference), significant transport occurs even at low amplitudes, indicating poor confinement. The transition between these regimes follows a distinct, curved boundary.

Panels (b) and (c) show the space $(\alpha_2, \alpha_1)$ for $\nu = 0$ and $\nu = \pi/2$, respectively. As expected, confinement degrades with increasing amplitude. Both spaces exhibit a well-defined boundary between no-transport and transport scenarios, but its shape is fundamentally different: linear for $\nu = 0$ and semi-circular for $\nu = \pi/2$. More generally, for equal spatial coupling $\eta_1=\eta_2$ the effective single-wave amplitude $A$ obtained in Appendix~\ref{Appendix3} satisfies
\begin{equation}\label{Eq20}
    \begin{split}
        &A^2 = (\alpha_1 + \alpha_2\cos\nu)^2 + (\alpha_2\sin\nu)^2 \\
&= \alpha_1^2 + \alpha_2^2 + 2\alpha_1\alpha_2\cos\nu.
    \end{split}
\end{equation}
Therefore the transmissivity/no-transmissivity boundary corresponding to a fixed critical amplitude $A_c$ is the locus
\begin{equation}\label{Eq21}
    \alpha_1^2 + \alpha_2^2 + 2\alpha_1\alpha_2\cos\nu = A_c^2,
\end{equation}
which is a quadratic curve in $(\alpha_1,\alpha_2)$ (a conic). For $|\cos\nu|<1$ this locus is an ellipse (restricted to the positive quadrant in our plots); it may appear as a semicircle/quarter-circle when $\cos\nu=0$), the special cases being $\nu=\pi/2$ (cos $\nu=0$) where one obtains $\alpha_1^2+\alpha_2^2=A_c^2$ (a circle) and $\nu=0$ (cos $\nu=1$) where the relation reduces to $(\alpha_1+\alpha_2)^2=A_c^2$ (a degenerate straight line).

This difference highlights the role of phase in shaping the interaction between waves, even with identical spatial modes.

In the next section, we will explore the more complex case of waves with different spatial modes.

\section{Distinct spatial-mode two-wave transport (mode-mismatched pairs)}\label{section5}
\begin{figure*}
    \centering
    \includegraphics[width=0.75\linewidth]{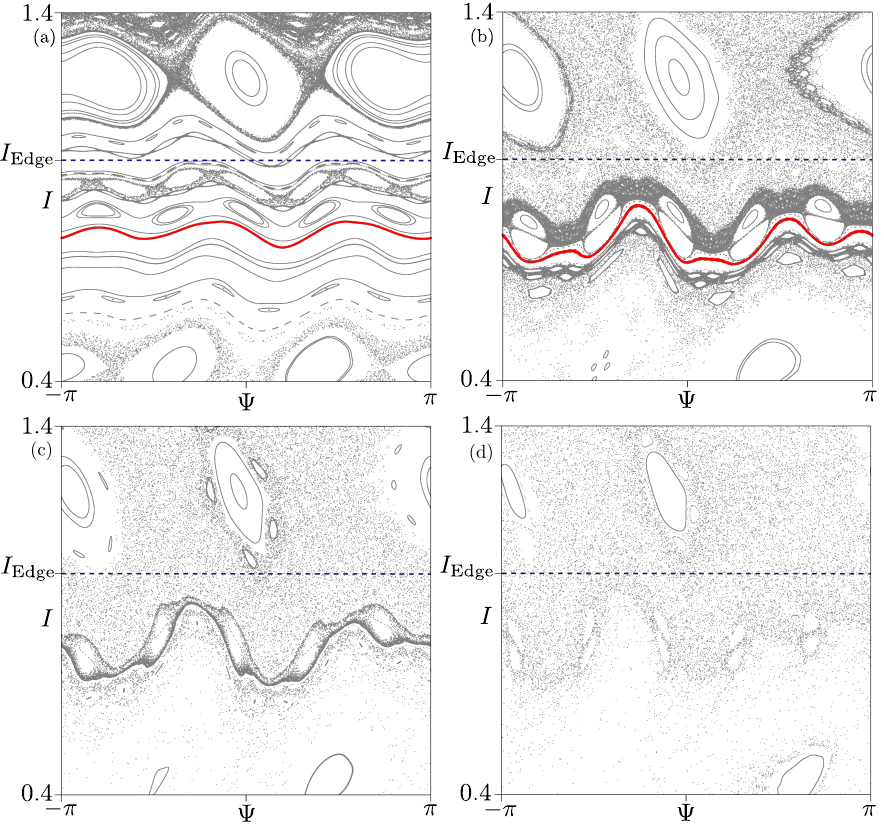}
    \caption{Impact of a secondary wave on phase space structure. Portraits in action-angle coordinates $(I, \Psi)$ for a fixed $\alpha_1 = 0.2$ with increasing amplitude of the second wave: (a) $\alpha_2 = 0.05$, (b) $\alpha_2 = 0.1$, (c) $\alpha_2 = 0.2$, (d) $\alpha_2 = 0.3$. The STB is highlighted in red in each panel.}
    \label{fig4}
\end{figure*}
\begin{figure*}
    \centering
    \includegraphics[width=0.85\linewidth]{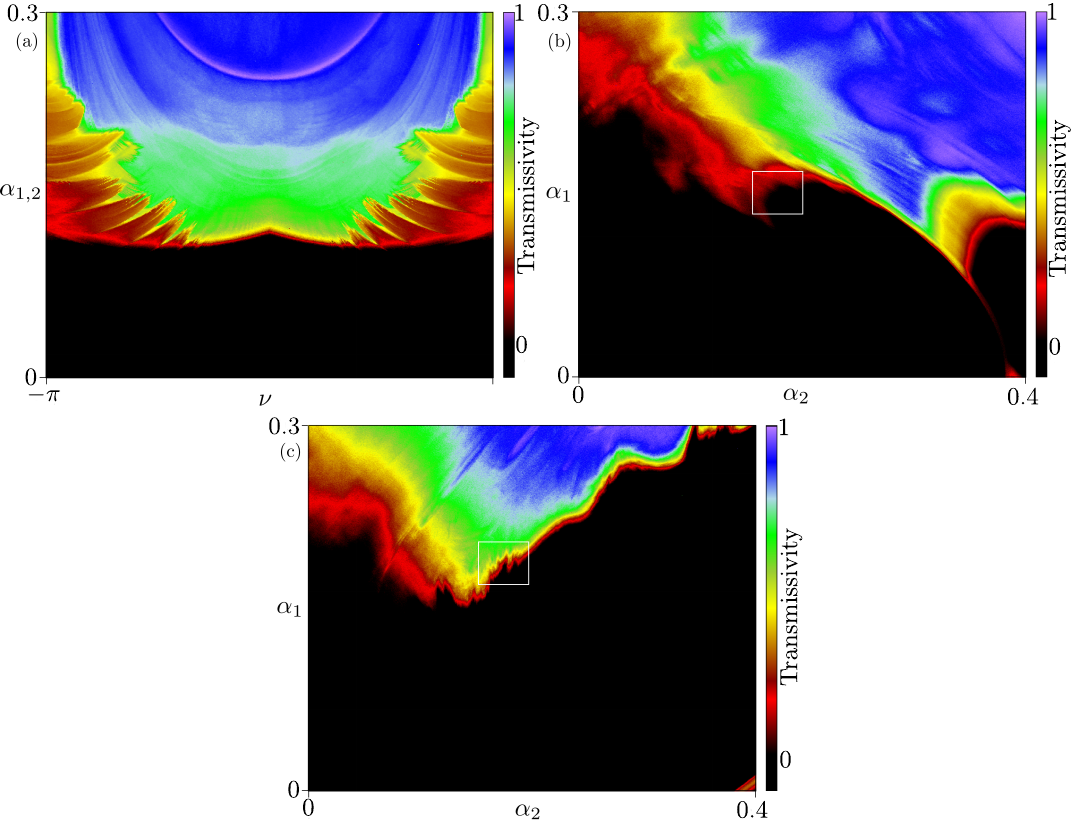}
    \caption{Transmissivity across different parameter spaces for $\eta_1 = 2$ and $\eta_2 = 3$. (a) Phase-amplitude space $(\nu, \alpha_{1,2})$. (b, c) Amplitude-amplitude space $(\alpha_2, \alpha_1)$ for a fixed phase difference of (b) $\nu = 0$ and (c) $\nu = \pi/2$. In all panels, black denotes parameter combinations resulting in no transport.}
    \label{fig5}
\end{figure*}

Having established the dynamics for identical spatial modes, we now introduce a difference by setting the modes to $\eta_1 = 2$ and $\eta_2 = 3$. This section focuses on mode-mismatched pairs; the choice $\eta_1=2,\eta_2=3$ is representative of the generic behavior that arises when the two waves do not share the same spatial harmonic. This allows us to examine the resulting changes to phase space dynamics and its consequence in transport.

Figure~\ref{fig4} presents this change through a sequence of phase space portraits (a-d) for fixed $\nu = 0$ and $\alpha_1 = 0.2$, with increasing amplitude $\alpha_2$ of the secondary wave. Even at a low amplitude ($\alpha_2 = 0.05$, panel a), the phase space exhibits a clear difference from the identical-mode case in Fig.~\ref{fig2}. The topology is significantly more complex, characterized by the emergence of secondary and higher-order resonances, in contrast to the predominant primary resonance structure in the identical-mode. A key feature is the presence of the STB (highlighted in red), indicating that its existence is not directly dependent on the specific spatial modes involved. However, as $\alpha_2$ increases from (a) to (d), its growing amplitude destroys regular orbits, giving rise to widespread chaos and a consequent significant increase in particle transport. We stress that similar qualitative behavior — earlier onset of resonances, richer higher-order resonance chains and sticky regions — is observed for other mismatched pairs $(\eta_1,\eta_2)$ (for example $\eta_1=1,\eta_2=3$ or $\eta_1=2,\eta_2=5$) provided the modes are different; the $\eta_1=2,\eta_2=3$ pair is presented here as a clear illustrative example.

To facilitate direct comparison with the identical-mode case, we characterize the transport dynamics of the distinct-mode system using the same transmissivity metric, exploring the $(\nu, \alpha_{1,2})$ and $(\alpha_2, \alpha_1)$ parameter spaces. Figure~\ref{fig5} presents the resulting transmissivity maps: (a) the phase-amplitude space $(\nu, \alpha_{1,2})$, and (b,c) the amplitude-amplitude space $(\alpha_2, \alpha_1)$ for phase differences $\nu = 0$ and $\nu = \pi/2$, respectively.

A comparison between Fig.~\ref{fig5}a and Fig.~\ref{fig3}a reveals significant differences, particularly in the boundary separating transport and no-transport regimes. For identical spatial modes, this boundary is defined by a smooth, well-behaved curve. In contrast, for distinct modes, the transition is not direct and exhibits a more complex structure. Furthermore, the distinct-mode case displays intricate fractal-like structures (visible in red/yellow), which correspond to stickiness phenomena in the Hamiltonian phase space. These structures, indicative of complex transport dynamics, are absent in the identical-mode scenario. This result underscores that the interference between waves with different spatial modes generates a richer and more complex transport landscape.

Overall, confinement is systematically reduced for mode-mismatched pairs. Two mechanisms explain this effect. 
First, the definition Eq.\,\eqref{Eq14} of the control parameter
\[
    \alpha_k \propto \eta_k\,\phi_k,
\]
implies that for comparable wave amplitudes \(\phi_k\), larger spatial wavenumbers \(\eta_k\) produce larger effective perturbation amplitudes \(\alpha_k\); increased \(\alpha\) weakens invariant tori and facilitates the onset of chaos.  Second, the difference in spatial harmonics leads to a more complex resonance-overlap structure: distinct \(\eta\) introduce additional periodicities and higher-order resonances that overlap with the primary resonance at lower perturbation values, accelerating the destruction of transport barriers and increasing stickiness. Together, these effects cause earlier destruction of the barriers and therefore reduced confinement for distinct \(\eta\) values.

The same comparative analysis holds for the amplitude parameter spaces in Figs.~\ref{fig3}(b-c) and~\ref{fig5}(b-c). In the identical-mode scenario, not only is the fundamental transport-no-transport boundary a well-defined curve, but the gradients in transmissivity (indicating the intensity of stickiness) also vary smoothly. In stark contrast, the distinct-mode case exhibits a complex and fragmented structure. The transitions are ill-defined, culminating in the fractal-like patterns that signify non-trivial interference effects and rich stickiness dynamics. In the following section, we introduce a numerical metric to quantitatively determine this increased complexity in the transport dynamics.

\section{Fractal-like boundary}\label{section6}
\begin{figure}
    \centering
    \includegraphics[width=\linewidth]{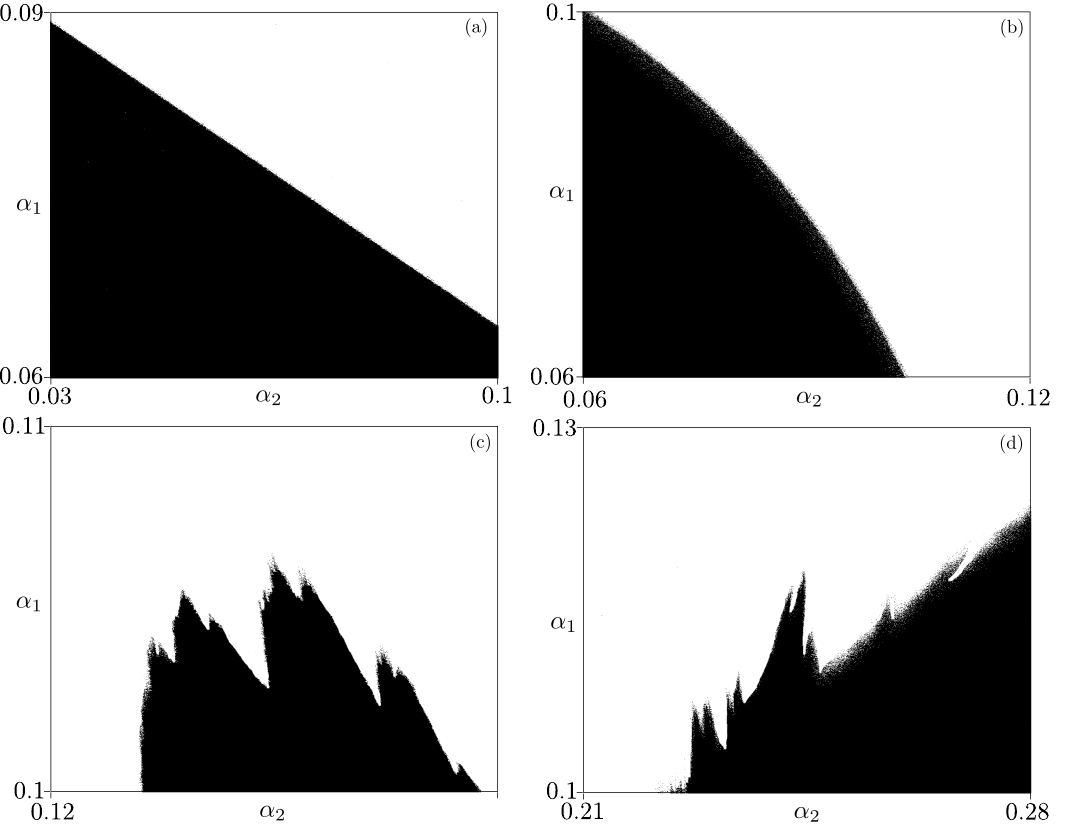}
    \caption{Magnified views of the transmissivity boundaries from selected regions in Figs.~\ref{fig3} and~\ref{fig5}. Panels (a-d) correspond to the areas within the white rectangles in (a) Fig.~\ref{fig3}b, (b) Fig.~\ref{fig3}c, (c) Fig.~\ref{fig5}b, and (d) Fig.~\ref{fig5}c, respectively. The plots compare boundary structures for different spatial mode configurations: (a,b) identical spatial modes ($\eta_{1,2} = 1$) at $\nu = 0$ and $\nu = \pi/2$; (c,d) different spatial modes ($\eta_1 = 2$, $\eta_2 = 3$) at the same phase values. Black denotes parameter combinations with zero transmissivity (no transport).}
    \label{fig6}
\end{figure}

The transmissivity parameter spaces presented thus far exhibit boundaries of varying geometric complexity, separating transport from no-transport regimes. These boundaries range from smooth curves to intricate, fractal-like structures. In order to quantitatively characterize this complexity, we employ the uncertainty exponent method \cite{grebogi1987multi} to compute the box-counting dimension of the boundaries.

The box-counting dimension $d$ provides a measure of a set's geometric complexity. For a boundary embedded in a $D$-dimensional parameter space, $d$ is defined by considering the minimum number of boxes $N(\delta)$ of side length $\delta$ needed to cover it:
\begin{equation}\label{Eq22}
d = \lim_{\delta \rightarrow 0} \dfrac{\ln N(\delta)}{\ln (1/\delta)}.
\end{equation}
This implies the scaling $N(\delta) \propto \delta^{-d}$ for small $\delta$. The uncertainty exponent $\beta$ is then given by $\beta = D - d$. It represents the scaling of the "uncertain" volume—the region of parameter space within distance $\sim\delta$ of the boundary—which scales as $\delta^\beta$. In our two-dimensional parameter spaces ($D=2$), a smooth boundary ($d=1$) corresponds to $\beta=1$, while a fractal boundary ($1 < d < 2$) yields $0 < \beta < 1$.

To estimate $d$ numerically, we implemented a routine that discretizes the parameter space into a grid of size $G \times G$, with $G$ ranging from $50$ to $200$. The transmissivity (classified as zero or non-zero) is used to assign each point to the transport or no-transport region. For a given grid size, a box is labeled "uncertain" if it contains points from both classes. The number of such boxes, $N(\delta)$, is recorded for the corresponding box size $\delta$ (accounting for the aspect ratio). The box-counting dimension $d$ is then obtained from the slope of a linear fit to $\ln N(\delta)$ versus $\ln(1/\delta)$. The final reported value is the mean of $d$ over all grid sizes, with the standard deviation providing the uncertainty.

We applied this analysis to four magnified regions of interest, marked by white rectangles in Figs.~\ref{fig3} and~\ref{fig5}. Figure~\ref{fig6} presents these close-ups: panels (a) and (b) correspond to the identical spatial mode configuration for $\nu = 0$ and $\nu = \pi/2$, respectively, while panels (c) and (d) show equivalent regions for the distinct spatial mode configuration. The computed dimensions are: (a) $1.05 \pm 0.02$, (b) $1.02 \pm 0.03$, (c) $1.42 \pm 0.03$, and (d) $1.35 \pm 0.06$.

These results confirm a fundamental difference in boundary structure: boundaries (a) and (b) are smooth ($d = 1$), whereas (c) and (d) are fractal ($1 > d > 2$). Notably, in case (a) the measured boundary dimension remains above the analytically expected value even when measurement uncertainties are taken into account, which indicates a limitation of the applied method. This quantifies the visual observation that wave interference with distinct spatial modes generates more intricate and unpredictable dynamics. This complexity is closely linked to  stickiness \cite{mugnaine2018, mugnaine2019, sales2023stickiness, lozej2020stickiness}, a phenomenon that strongly influences transport. Stickiness is highly sensitive to parameters ($\nu, \alpha_{1,2}$) and iteration count, making transport dynamics difficult to predict and directly contributing to the observed fractal boundaries. Consequently, stickiness plays a significant role in particle trajectories near the fractal boundaries (c) and (d). In contrast, the smooth boundaries (a) and (b) suggest stickiness has a markedly reduced, though non-negligible, effect.

\section{Conclusions}\label{conclusion}

Modeling the electrostatic potential as a superposition of waves whose spatial modes are integer multiples with phase offsets, we derive a two-dimensional symplectic multiple wave map that captures particle dynamics in the plasma edge. To enable a systematic study of how perturbation amplitude and relative phase affect transport, we reduce the general map to a two-wave model and analyze representative combinations of spatial modes.

For two waves with identical spatial modes, the phase difference $\nu$ plays a decisive role in confinement. When $\nu=\pm\pi$, destructive interference cancels the wave drive and the system becomes effectively integrable, yielding strong confinement. By contrast, when $\nu=0$ the waves add constructively, producing a large perturbation that destroys regular orbits and enhances chaotic transport, thereby degrading confinement. Using transmissivity as our confinement criterion, we find smooth boundaries between no-transport and transport regimes, consistent with a direct interplay between the two waves in this case.

When the two waves have distinct spatial modes, the interaction is less direct and the phase space becomes considerably more complex: higher-order resonances and sticky regions appear and dominate the dynamics. The transmissivity parameter spaces reflect this complexity, exhibiting intricate boundary geometry between no-transport and transport scenarios. Overall, our results show that both relative phase and the choice of spatial modes critically determine whether wave superposition enhances confinement or promotes chaotic transport.

To objectively evaluate this increased complexity, we computed the box-counting dimension of these boundaries. The results provide definitive quantitative evidence: for identical modes, the dimension is approximately $1$, confirming a smooth curve. For distinct modes, the dimension is fractal (non-integer), revealing a complex geometric structure. This finding conclusively links the observed complex transport dynamics to the underlying fractal-like geometry of the stability boundary in parameter space. 

We stress that this study addresses the transport consequences of prescribed phase relations and spatial spectra within a reduced Hamiltonian test-particle model. In realistic plasmas, modulational and other nonlinear wave–wave processes may reorganize spectral energy and induce phase correlations, potentially modulating the transport mechanisms discussed here. Therefore, extending the present particle-dynamics approach by coupling it to self-consistent modulational, envelope, or kinetic–wave models constitutes an important direction for future work.

In summary, our results demonstrate that both the relative phase and the spatial spectrum of interacting waves are important to the particle transport that affect the plasma confinement. Wave interference can be adjusted to either suppress transport through destructive interference or to generate a complex mix of regular and chaotic/sticky dynamics through mode mismatch. These insights underscore the importance of considering the full spectral nature of turbulence in fusion plasmas to accurately predict and control confinement properties.

\begin{acknowledgments}
The authors acknowledge financial support from the Brazilian funding agencies: Conselho Nacional de Desenvolvimento Científico e Tecnológico (CNPq) under Grant Nos. 402123/2024-7, 302665/2017-0, and 200423/2025-8; Fundação de Amparo à Pesquisa do Estado de São Paulo (FAPESP) under Grant Nos. 2024/05700-5 and 2019/07329-4; and Coordenação de Aperfeiçoamento de Pessoal de Nível Superior (CAPES) under Grant No. 88887.843528/2023-00. 

The authors also thank the members of the Grupo de Controle de Oscilações, Instituto de Física, Universidade de São Paulo (USP), and of the PIIM Laboratory for helpful discussions and support.
\end{acknowledgments}

\section*{Data Availability}
The data that support the findings of this article are openly
available \cite{SouzaDataset2026}.

\appendix

\section{Parameters Normalization}\label{Appendix1}

The normalization applied involves dividing the parameters of the problem $(E_r, B, a, \omega_0, R, \text{ and } v_{||})$ by factors that have the same dimensions and units as these quantities, aiming to bring their typical values close to one. The minor radius of the plasma, $a$, is normalized by a factor $a_0$ with the dimension of length, representing the characteristic length scale of the system. This characteristic length was chosen as the minor radius of the plasma itself, $a_0 = 0.18$\,m, such that $a' = a/a_0 = 1$ in the new scale. Similarly, we chose $B_0 = 1.1$\,T, which gives $B' = B/B_0 = 1$. The normalization factor for the electric field was chosen such that $|E_r(a)| = 1$, i.e., $E_0 = |E_r(a)|$. Thus,

\begin{align*}
    E_{r}^{'} &= \dfrac{E_r}{E_0},   \quad   \phi_{k}^{'} = \dfrac{\phi_{k}}{aE_0}, \quad     v_{||}^{'}=\dfrac{B}{E_0}v_{||}    \\
    t^{'} &= \dfrac{E_0}{aB}t,  \quad   \omega_{0}^{'} = \dfrac{aB}{E_0}\omega_0
\end{align*}

\section{Derivation of the Floating Potential Expression}\label{Appendix2}

Starting from the electrostatic potential, Eq.\,\eqref{Eq5}, formed by a superposition of infinitely many waves with multiple integer spatial modes and phase differences 
\begin{equation}\tag{\ref{Eq5}}
 \tilde{\phi}(\Psi, t) =  \sum_k \sum_{n = -\infty}^{\infty} \phi_{n}^{k} \cos (\eta_k \Psi - n\omega_0 t + \nu_{n}^{k}),
\end{equation}
we apply the trigonometric identity
\begin{equation}\label{B1}
\begin{split}
    &\cos (\eta_k\Psi - n\omega t + \nu_{n}^{k}) = \\
    &\cos (\eta_k\Psi +\nu_{n}^{k} )\cos  n\omega t + \sin(\eta_k\Psi + \nu_{n}^{k} )\sin  n\omega t.
\end{split}
\end{equation}
Substituting Eq. \eqref{B1} into Eq. \eqref{Eq5}, the potential becomes:
\begin{equation}\label{B2}
\begin{split}
    \tilde{\phi}(\Psi, t) = \sum_k \sum_{n = -\infty}^{\infty}\phi_{n}^{k} \{&  \cos (\eta_k\Psi +\nu_{n}^{k} )\cos  n\omega t \\
    &+ \sin(\eta_k\Psi + \nu_{n}^{k} )\sin  n\omega t \}.
\end{split}
\end{equation}
The Poisson summation formula converts the infinite series into Dirac delta distribution. Noting that $\sum_{n=-\infty}^{\infty} \sin(n\omega t) = 0$ due to symmetry, we retain only the cosine terms:
\begin{subequations}
\begin{align}
\sum_{n= -\infty}^{\infty} \cos (n\omega t) &= 2\pi \sum_{n = -\infty}^{\infty} \delta (\omega t - 2\pi n), \label{B7a} \\
\sum_{n = -\infty}^{\infty} \sin (n\omega t) &= 0. \label{B7b}
\end{align}
\end{subequations}
for $v_{n}^{k} = v_k$ only. Substituting these into the expression for \(\tilde{\phi}(\Psi, t)\), the final form of the potential becomes:
\begin{equation}\tag{\ref{Eq6}}
\tilde{\phi}(\Psi, t) =   2\pi \sum_k \phi_{k} \cos(\eta_k \Psi + \nu_{k})\sum_n \delta(\omega_0 t - 2\pi n).
\end{equation}

\section{Derivation of the reduced map}\label{Appendix3}
We consider an electrostatic potential \(\tilde{\phi}\) formed by the superposition of two waves with identical spatial modes but a relative phase difference \(\nu\). The potential is expressed as:
\begin{equation}
\begin{split}
    \tilde{\phi} &= \tilde{\phi}_{1} + \tilde{\phi}_{2}, \\
    &= \sum_{n = -\infty}^{\infty} \left[\phi_{1} \cos (\Psi_n - n\omega_0 t) + \phi_{2} \cos (\Psi_n - n\omega_0 t + \nu) \right].
\end{split}
\end{equation}

This expression can be reformulated using Dirac delta distributions to capture the impulsive nature of the interaction:
\begin{equation}
    \tilde{\phi} = 2\pi \sum_{n = -\infty}^{\infty} \left(C \cos \Psi_n - D \sin \Psi_n\right)\delta (\omega_0 t - 2\pi n),
\end{equation}
where the coefficients \(C\) and \(D\) incorporate the phase difference:
\begin{equation}
    C = \phi_1 + \phi_2 \cos \nu, \quad D = \phi_2 \sin \nu.
\end{equation}

From this potential, we derive the following drift wave map:
\begin{subequations}
\begin{align}
    I_{n+1} &= I_n + C\sin \Psi_n + D\cos \Psi_n, \\
    \Psi_{n+1} &= \Psi_n + g(I_{n+1}).
\end{align}
\end{subequations}

The two trigonometric terms in the action update can be combined into a single sinusoidal function with a phase shift. Using the trigonometric identity:
\begin{equation}
    \sin(\Psi + \zeta) = \sin\Psi \cos\zeta + \cos\Psi \sin\zeta,
\end{equation}
we seek to express \(C\sin\Psi + D\cos\Psi\) as \(A\sin(\Psi + \zeta)\). Multiplying the identity by \(A\):
\begin{equation}
    A\sin(\Psi + \zeta) = A\sin\Psi \cos\zeta + A\cos\Psi \sin\zeta,
\end{equation}
and matching coefficients with \(C\sin\Psi + D\cos\Psi\) yields:
\begin{subequations}
\begin{align}
    A\cos\zeta &= C, \label{eq:C7a} \\
    A\sin\zeta &= D. \label{eq:C7b}
\end{align}
\end{subequations}

Squaring and adding Eqs.~\eqref{eq:C7a} and \eqref{eq:C7b} gives the amplitude:
\begin{equation}
    A^2(\cos^2\zeta + \sin^2\zeta) = C^2 + D^2 \implies A = \sqrt{C^2 + D^2}.
\end{equation}
The phase shift \(\zeta\) is obtained by dividing Eq.~\eqref{eq:C7b} by Eq.~\eqref{eq:C7a}:
\begin{equation}
    \tan\zeta = \frac{D}{C} \implies \zeta = \tan^{-1}\left( \frac{D}{C} \right).
\end{equation}

Thus, the combined expression becomes:
\begin{equation}
    C \sin\Psi + D \cos\Psi = A\sin(\Psi + \zeta).
\end{equation}

Defining a shifted angle variable \(\theta = \Psi + \zeta\), the map simplifies to:
\begin{align}
    I_{n+1} &= I_n + A\sin\theta_n, \\
    \theta_{n+1} &= \theta_n + g(I_{n+1}).
\end{align}

This reduction consolidates the effect of two waves with identical spatial modes into a single effective wave with amplitude \(A\) and phase shift \(\zeta\), significantly simplifying the analysis of phase-dependent transport phenomena.

\bibliography{main}% Produces the bibliography via BibTeX.

\end{document}